\documentclass[12pt]{article}
\usepackage{sc3conf}
\usepackage{amsfonts}
\usepackage{epsfig}

\def\a{{\alpha}}
\def\b{{\beta}}
\def\d{{\delta}}
\def\e{{\epsilon}}
\def\g{{\gamma}}

\def\s{{\sigma}}
\def\G{{\Gamma}}
\def\CD{{\cal D}}

\def\CL{{\cal L}}

\def\CN{{\cal N}}
\def\CO{{\cal O}}
\def\IB{\relax{\rm I\kern-.18em B}}
\def\IC{{\relax\hbox{$\inbar\kern-.3em{\rm C}$}}}
\def\ID{\relax{\rm I\kern-.18em D}}
\def\IE{\relax{\rm I\kern-.18em E}}
\def\IF{\relax{\rm I\kern-.18em F}}
\def\IG{\relax\hbox{$\inbar\kern-.3em{\rm G}$}}
\def\IGa{\relax\hbox{${\rm I}\kern-.18em\Gamma$}}
\def\IH{\relax{\rm I\kern-.18em H}}
\def\II{\relax{\rm I\kern-.18em I}}
\def\IK{\relax{\rm I\kern-.18em K}}
\def\IP{\relax{\rm I\kern-.18em P}}
\def\IQ{\relax\hbox{$\inbar\kern-.3em{\rm Q}$}}
\def\IR{\relax{\rm I\kern-.18em R}}

\def\IZ{{\rm Z}\kern-3.8pt {\rm Z} \kern2pt}
\def\hG{{\hat\Gamma}}
\def\half {{1\over 2}}
\def\beq{\begin{equation}}
\def\eeq{\end{equation}}
\def\beqa{\begin{eqnarray}}
\def\eeqa{\end{eqnarray}}
\def\p{\partial}

\def\tw{{\tilde w}}
\def\tS{{\tilde S}}
\def\tL{{\tilde L}}

\begin{document}
\raggedbottom

\title{Large N dualities from wrapped D--branes}

\authors{Jos\'e D. Edelstein}

\addresses{Departamento de Matem\'atica, Instituto Superior Tecnico \\
Av. Rovisco Pais, 1049--001, Lisboa, Portugal
  \nextaddress Instituto de F\'\i sica de La Plata -- Conicet,
Universidad Nacional de La Plata \\
C.C. 67, (1900) La Plata, Argentina \nextaddress Departamento de
F\'\i sica de Part\'\i culas, Universidad de Santiago de
Compostela \\ E-15782 Santiago de Compostela, Spain}

\maketitle

\begin{abstract}
We review some aspects of the gravity duals of supersymmetric
gauge theories, arising in the world--volume of D--branes wrapping
supersymmetric cycles of special holonomy manifolds, within the
framework of lower dimensional gauged supergravity.
\end{abstract}

\medskip
\medskip

In a seminal paper, 't Hooft proposed that large $N$ gauge
theories, in the strongly coupled regime, should be better
described by perturbative closed strings \cite{thooft}. The
argument goes as follows. Consider a gauge theory with the gauge
field $A_\mu^{k\bar k} \in F \otimes \bar F$ in the adjoint
representation of $U(N)$ and gauge coupling $\lambda$. Each line
in the Feynman diagrams becomes a double line and the diagram
itself is then drawn as a ribbon graph. These graphs can be
topologically classified according to the closed Riemann surface
$\Sigma_g$ of genus $g$ on which they can be drawn. In presence of
matter (that we will not consider, for simplicity, from now on),
the surface will also display holes. Any amplitude can then be
written as a sum over topologies
\begin{equation}
\label{amplit}
{\cal A} = \sum_{g=0}^\infty c_g (t) N^{2 - 2g} ~,
\end{equation}
where $t \equiv \lambda^2 N$ is the so--called 't Hooft parameter.
When $N \to \infty$ with $t$ fixed (the 't Hooft limit), planar
Feynman diagrams dominate and we have two well distinct
regimes: If $t << 1$, also $\lambda << 1$, and the gauge theory is
well described perturbatively. On the other hand, when $t >> 1$,
the amplitudes (\ref{amplit}) can be rearranged as
\begin{equation}
\label{amplit2}
{\cal A} = \sum_{g=0}^\infty \lambda_s^{2g - 2} {\cal A}_g (t) ~,
\end{equation}
where ${\cal A}_g (t)$ is a closed string amplitude on $\Sigma_g$,
$\lambda_s \equiv \lambda^2$ being the corresponding string
coupling, and $t$ is a {\em modulus} of the target space.

Early examples of 't Hooft's duality, involving bosonic strings on
various backgrounds that are dual to zero dimensional gauge
theories, were constructed some ten years ago \cite{kdv}. In the
same vein, it was recently proposed that type IIB superstring on
$AdS_5 \times S^5$ is dual to ${\cal N} = 4$ super Yang--Mills
theory in four dimensions \cite{malda}. The gauge theory is
realized on the world--volume of D3--branes. This conjecture was
extended to theories with sixteen supercharges that correspond to
the low--energy dynamics of {\em flat} D--branes. These are, in
general, non--conformal, and the gravity/gauge theory
correspondence provides a powerful tool to study the phase
structure of the resulting RG flows \cite{imsy}.

Analogue results can be obtained in the context of topological
strings. The A-model topological string on the resolved conifold
is dual to Chern--Simons gauge theory on $S^3$ \cite{gova}. There
is also a mirrored version of this: B--model topological strings
on local Calabi--Yau threefolds being dual to matrix models
\cite{dijva}. The embedding of these dualities into superstring
theory allowed Vafa to conjecture that ${\cal N} = 1$ super
Yang--Mills theory in four dimensions must be dual to either type
IIA superstrings on $\CO(-1) + \CO(-1) \to \IP^1$ with RR fluxes
through the exceptional $\IP^1$ or, through the looking glass, to
type IIB superstrings on the deformed conifold with RR fluxes
piercing the blown--up $S^3$ \cite{vafa}. In these cases, the
gauge theory is realized on the world--volume of {\em wrapped}
D--branes in a Calabi--Yau threefold; respectively, D6--branes on
special Lagrangian three--cycles and D5--branes on holomorphic
two--cycles. Furthermore, arbitrary tree level superpotentials can
be accommodated into this framework \cite{civeot}.

The low--energy dynamics of a collection of D--branes wrapping
supersymmetric cycles is governed, when the size of the cycle is
taken to zero, by a lower dimensional supersymmetric gauge theory
with less than sixteen supercharges \cite{mn}. The non--trivial
geometry of the world--volume leads to a gauge theory in which
supersymmetry is appropriately twisted \cite{bvs}. The amount of
supersymmetry preserved has to do with the way in which the cycle
is embedded in a higher dimensional space. When the number of
branes is taken to be large, the near horizon limit of the
corresponding supergravity solution provides a gravity dual of the
field theory arising in their world--volume. The gravitational
description of the strong coupling regime of these gauge theories
allows for a geometrical approach to the study of such important
aspects of their infrared dynamics as, for example, chiral
symmetry breaking, gaugino condensation, domain walls, confinement
and the existence of a mass gap.

A natural framework to perform the above mentioned twisting is
given by lower dimensional gauged supergravities. Their domain
wall like vacuum solutions usually correspond to the near horizon
limit of D--brane configurations \cite{bst} thus giving directly
the gravity dual description of the gauge theories living on their
world--volumes. Let us consider, in this talk, the case of the
D6--brane. This system is best described in the infrared by means
of $\CN = 2$ seven dimensional super Yang--Mills theory
\cite{imsy}. So, for example, wrapping D6--branes on $S^3$ would
imply, after appropriate twisting, breaking one quarter of the
supercharges, the theory reducing to pure $\CN = 1$ four
dimensional super Yang--Mills in the infrared. The above referred
twisting corresponds to $S^3$ being a special Lagrangian
submanifold of a Calabi--Yau three--fold. The natural set up for
this problem is provided by eight dimensional gauged supergravity
\cite{ss}. The Lagrangian describing the dynamics of a sector of
the theory given by the metric $g_{\mu\nu}$, the dilaton $\Phi$,
five scalars $L_\a^i$ in the coset $SL(3,\IR)/SO(3)$, and an
$SU(2)$ gauge potential $A^i$, reads \beq \label{boslag} e^{-1}
\CL = {1 \over 4} R - {1 \over 4} e^{2\Phi} (F_{\mu\nu}^{~i})^2 -
{1 \over 4} (P_{\mu ij})^2 - \half (\p_\mu\Phi)^2 - {1 \over 16}
e^{-2\Phi} (T_{ij} T^{ij} - \half T^2) ~, \eeq where $e$ is the
determinant of the vierbein $e_\mu^a$, $F_{\mu\nu}^i$ is the
Yang--Mills field strength and $P_{\mu ij}$ is a symmetric and
traceless quantity defined by \beq \label{pyq} P_{\mu ij} + Q_{\mu
ij} \equiv L_i^\a (\p_\mu \d_\a^{~\b} - \e_{\a\b\g} A_\mu^\g)
L_{\b j} ~, \eeq $Q_{\mu ij}$ being the antisymmetric counterpart
and \beq \label{ttensor} T^{ij} \equiv L_\a^i L_\b^j \d^{\a\b} ~,
~~~~~~~ T = \d_{ij} T^{ij} ~. \eeq The supersymmetry
transformations for the fermions are given by \beq \label{susytr1}
\d\psi_\g = \CD_\g \e + {1 \over 24} e^{\Phi} F_{\mu\nu}^i \hG_i
(\G_\g^{~\mu\nu} - 10 \d_\g^{~\mu} \G^\nu) \e - {1 \over 288}
e^{-\Phi} \e_{ijk} \hG^{ijk} \G_\g T \e ~, ~~~~~~~ \eeq \beq
\label{susytr2} \d\chi_i = \half (P_{\mu ij} + {2 \over 3} \d_{ij}
\p_\mu\Phi) \hG^j \G^\mu \e - {1 \over 4} e^{\Phi} F_{\mu\nu i}
\G^{\mu\nu} \e - {1 \over 8} e^{-\Phi} (T_{ij} - \half \d_{ij} T)
\e^{jkl} \hG_{kl} \e ~, \eeq where we use, for the Clifford
algebra, $\G^a = \g^a \times \II$, $\hG^i = \g_9 \times \s^i$,
$\g^a$ are eight dimensional gamma matrices, $\g_9 = i \g^0 \g^1
\dots \g^7$, $\g_9^2 = 1$, and $\s^i$ are the Pauli matrices. It
is convenient to introduce $\hG_9 \equiv -i \hG^{123} = \g_9
\times \II$.

Let us start by considering a flat D6--brane configuration in
which we only excite the dilaton and one scalar $\varphi$, and the
ansatz for the line element is \beq \label{flatds} ds^2 = e^{2
f(\rho)} dx^2_{1,6} + d\rho^2 ~. \eeq The corresponding BPS
equations, emerging from $\d\psi_\g = \d\chi_i = 0$, are \beq
\label{bpsfl1} \Phi'(\rho) = {1 \over 8} e^{-\Phi} (e^{-4\varphi}
+ 2 e^{2\varphi}) ~, \eeq \beq \label{bpsfl2} \varphi'(\rho) = {1
\over 6} e^{-\Phi} (e^{-4\varphi} - e^{2\varphi}) ~, \eeq while
the equations for $f$ and $\e$ can be easily integrated with the
result \beq \label{flatsol} f = {1 \over 3} \Phi ~, ~~~~~~~~ \e =
i \hG_9 \G_r \e = e^{{1 \over 6} \Phi} \e_0 ~. \eeq After the
change of variables $d\rho = e^{\Phi - 2\varphi} dt$, the BPS
equations decouple and we obtain the solution \footnote{Notice
that the solution presented in \cite{bst,en} corresponds to $\xi_0
= 0$.} \beq \label{intsol} \varphi(t) = {1 \over 6} \left[ \log
(e^t - \xi_0) - t \right] ~, ~~~~~ \Phi(t) = {3 \over 4} \left[
\varphi(t) + \half (t - t_0) \right] ~. \eeq A further change of
variables $e^t = r^4 - a^4$, with $\xi_0 = - a^4$, drives the
solution, when uplifted to eleven dimensions through the external
$S^3$ (whose left invariant Maurer--Cartan one forms we denote as
$\tw^i$), to the form \beq \label{eghan} ds_{11d}^2 = dx^2_{1,6} +
{1 \over {1 - {a^4 \over r^4}}} dr^2 + {r^2 \over 4} \left[
(\tw^1)^2 + (\tw^2)^2 + \left( {1 - {a^4 \over r^4}} \right)
(\tw^3)^2 \right] ~, \eeq where, besides the seven dimensional
Minkowskian contribution from the uplift of the world--volume of
the flat D6--branes, we get a (hyperk\"ahler) metric for a
non--trivial ALE four manifold with $SU(2) \times U(1)$ isometry,
namely the Eguchi--Hanson metric \cite{eh}. This is in coincidence
with the uplifting of the near horizon solution corresponding to
D6--branes in type IIA. It is natural to analyze these
configurations in 11d for the fact that uplifted D6 branes become
purely gravitational. Besides, the D6 branes are strongly coupled
in the ultraviolet and the would be decoupling limit has to be
addressed in eleven dimensions. In particular, the 11d
supergravity solution is trustable for any number of branes
\cite{imsy}.

We now proceed towards the supergravity dual of $\CN = 1$ super
Yang--Mills theory in four dimensions, arising in the low--energy
dynamics of D6--branes wrapped on a special Lagrangian $S^3$ in a
Calabi--Yau three--fold. Let us start with an ansatz that
describes such a deformation of the world--volume of the
D6--branes \beq \label{metrica} ds^2 = e^{2f(r)} dx^2_{1,3}+ {1
\over 4} e^{2h(r)} \sum_{i=1}^3 (w^i)^2 + dr^2 ~, \eeq where $w^i$
are the left invariant one forms corresponding to the special
Lagrangian $S^3$. The fields on the D6--branes transform under
$SO(1,6) \times SO(3)_R$ as ({\bf 8},{\bf 2}) for the fermions and
({\bf 1},{\bf 3}) for the scalars, while the gauge field is a
singlet under $R$--symmetry. When we wrap the D6 branes on a
three--cycle, the symmetry group splits as $SO(1,3) \times SO(3)
\times SO(3)_R$. The effect of the twisting is to preserve those
fields that are singlets under a diagonal $SO(3)_D$ build up from
the last two factors. The gauged $R$--symmetry is used to cancel
the effect of the spin connection in the covariant derivative
\cite{ens}. The vector fields survive but the scalars are
transformed into one forms on the curved surface, so we are left
with a theory with no scalar fields in the infrared; besides four
supercharges are preserved.

The twisting might be achieved by turning on the non--Abelian
$SO(3)$ gauge field given by the left invariant form of the three
sphere, \beq \label{amu} A^i = - {1 \over 2} ~w^i ~, \eeq and it
is easy to see that in this case we can get rid of the scalars
$L_\a^i = \d_\a^i ~ \Rightarrow ~ P_{ij} = 0 ~, ~Q_{ij} = -
\e_{ijk} A^k$. We impose the following projections in the
supersymmetric parameter $\e$: \beq \label{proju}
\Gamma_{ij}\,\e\,=\,-\hat\Gamma_{ij}\,\e\, ~,~~ i \neq j = 1, 2, 3
~~~~~~~ \e = i \hG_9 \G_r \e ~. \eeq These projections leave
unbroken $1/8$ of the original supersymmetries, that is, four
supercharges. The first order BPS equations are, \beq \label{buu}
f'(r) = {1\over 3} \Phi'(r) = - {1 \over 2} e^{\Phi - 2h} + {1
\over 8} e^{-\Phi} ~, \eeq \beq \label{bud} h'(r) = {3 \over 2}
e^{\Phi - 2h} + {1 \over 8} e^{-\Phi} ~, \eeq and the solution
\cite{en}, when uplifted to eleven dimensions, read: \beq
\label{met} ds^2 = dx^2_{1,3} + {1 \over \Bigl( 1- {a^3 \over
\rho^3} \Bigr)} d\rho^2 + {\rho^2 \over 12} (\tw^a)^2 + {\rho^2
\over 9} \Bigl( 1 - {a^3 \over \rho^3} \Bigr) \Bigl[ w^a - {1
\over 2} \tw^a \Bigr]^2 ~. \eeq This is the metric of a $G_2$
holonomy manifold which is topologically $\IR^4 \times S^3$. The
radial variable $\rho \geq a$ fills $S^3$ while the other sphere
$\tS^3$ remains of finite volume $a^3$ when the former shrinks.
The $G_2$ holonomy manifold has isometry group $SU(2)_L \times
SU(2)_{\tilde L} \times SU(2)_D$, the first two factors
corresponding to the left action on $S^3$ and $\tilde S^3$
respectively, and the last one is the diagonal subgroup of
$SU(2)_R \times SU(2)_{\tilde R}$. There is a flop transition in
which the two spheres are exchanged. In this case, M--theory
smooths out the singularity thanks to the existence of $C$--field
fluxes through the three--sphere.

There are two very different quotients of this manifold: a
singular one by modding out by $\IZ_N \subset U(1) \subset
SU(2)_L$, and a non--singular quotient if one instead chooses
$\IZ_N \subset U(1) \subset SU(2)_{\tilde L}$. This is due to the
fact that $S^3$ shrinks to a point when $\rho \to a$ while $\tilde
S^3$ has radius $a$. Modding out by $\IZ_N \subset U(1) \subset
SU(2)_L$ results in an $A_{N-1}$ singularity fibered over $\tS^3$
so that, after KK reduction along the circle corresponding to the
$U(1)$, one ends with $N$ D6--branes wrapped on a special
Lagrangian $\tS^3$ in a Calabi--Yau three--fold. The second case,
amounts to modding out by $\IZ_N \subset U(1) \subset SU(2)_\tL$,
which has no fixed points so the quotient a smooth manifold
admitting no normalizable supergravity zero modes. Thus, M--theory
on the latter has no massless fields localized in the
transverse four-dimensional spacetime. By a smooth interpolation
between these manifolds, M--theory realizes the mass gap of ${\cal
N}=1$ supersymmetric four-dimensional gauge theory
\cite{ach1,amv}. After KK reduction of the smooth manifold one
ends with a non--singular type IIA configurations (without
D6--branes) on a space with the topology of $\CO(-1) + \CO(-1) \to
\IP^1$ \cite{amv}, and with $N$ units of RR flux through the
finite radius $S^2$.

Let us now consider turning on some units of RR four--form flux
along the unwrapped directions \cite{epr1}. The bosonic truncation
of eight dimensional supergravity relevant for our purposes now
includes a three-form potential (whose field strength we will
denote by $G$). This, in general, is an inconsistent truncation:
$G$ acts as a non--linear source for some of the forms we have
turned off. However, we will consider solutions that are fully
compatible with the equations of motion of 8d gauged supergravity
by imposing $G \wedge G = \ast G \wedge F^i = 0$, where $F^i$ is
the $SU(2)$ field strength and $\ast G$ is the Hodge dual of $G$
in eight dimensions. The presence of this flux introduces a
distinction between one of the unwrapped directions of the brane
and the other three. Accordingly, the ansatz for the metric will
be: \beq ds^2_8\,=\,e^{2f}\,dx_{1,2}^2\,+\,e^{2\alpha}\,dy^2\,+\,
{1 \over 4} ~e^{2h}\, \sum_{i=1}^3 (w^i)^2\,+\,dr^2\,\,,
\label{uno} \eeq where $f$, $\alpha$ and $h$ are functions of the
radial coordinate $r$. The corresponding ansatz for the 4-form $G$
in flat coordinates is $G_{012r} = \Lambda\, e^{-\alpha-3h-2\phi}$
with $\Lambda$ being a constant and $\phi$ the eight-dimensional
dilaton. The non--Abelian gauge field $A^i$ is chosen as in
(\ref{amu}) to undertake the prescribed twisting.

The supersymmetry transformations include now the contribution of
the $G$ field. The standard projections corresponding to the
D6--branes wrapping the $S^3$ (\ref{proju}) are supplemented by a
new one due to the presence of the $G$ flux. In flat indices,
$\Gamma_{012}\,\e\,=\,\e$. The number of supercharges unbroken by
this configuration is then one half of those corresponding to the
case $\Lambda = 0$, {\em i.e.} two. The BPS equations are: \beqa
f'&=&-{1\over 2}\,e^{\phi-2h}\,+\,{1 \over
8}\,e^{-\phi}\,+\,{\Lambda\over 2}\, e^{-\phi-3h-\alpha}\,\,,\\
\alpha'&=&-{1\over 2}\,e^{\phi-2h}\,+\,{1 \over
8}\,e^{-\phi}\,-\,{\Lambda\over 2}\, e^{-\phi-3h-\alpha}\,\,,\\
h'&=&{3\over 2}\,e^{\phi-2h}\,+\,{1 \over
8}\,e^{-\phi}\,-\,{\Lambda\over 2}\, e^{-\phi-3h-\alpha}\,\,,\\
\phi'&=&-{3\over 2}\,e^{\phi-2h}\,+\,{3\over
8}\,e^{-\phi}\,-\,{\Lambda\over 2}\, e^{-\phi-3h-\alpha}\,\,.
\label{catorce} \eeqa The general solution of this system can be
found \cite{epr1}, and its 11d uplift results into the following
metric: \beq ds^2_{11}\,=\, \big[\,H(\rho)\,\big]^{-{2\over
3}}\,dx_{1,2}^2\,+\, \big[\,H(\rho)\,\big]^{{1\over 3}}\,\Big[\,
dy^2\,+\,ds^2_7\,\Big]\,\,, \label{cicuatro} \eeq where $ds^2_7$
is the $G_2$ holonomy metric (\ref{met}), while the warp factor
$H(\rho)$ is, $$ H(\rho)\,=\,1+\, {1296 \over 5} \sqrt{3} {\Lambda
\over (12)^{1 \over 6}} \,\Bigg[\,{5\over 3 a^3\rho^2}\, {1\over
1\,-\,{a^3\over \rho^3}}+{10\over 3\sqrt{3}\,a^5}\, {\rm
arccot}\,\big[\,{2\rho +a\over a\sqrt{3}}\big] $$ \beq -\,{5\over
9a^5}\,\log\big(\,1\,+\,{3 a\rho\over (\rho-a)^2} \big)\,
\Bigg]\,, \label{citres} \eeq the four--form being given by
$F_{012\rho} = \e_{012} \partial_\rho [H(\rho)]^{-1}$. This
solution represents a smeared distribution of M2--branes on the
manifold of $G_2$ holonomy obtained before. $H(\rho)$ is an
harmonic function in the seven--manifold.

The somehow unusual appearance of a smeared configuration in this
approach deserves some comments. We should first remind that, even
in the case of flat D--branes, it is well known that D2--branes
have a low energy range, $g^2_{YM} < U < g^2_{YM} N^{1 \over 5}$,
in which string theory is strongly coupled but the eleven
dimensional curvature is small, and the appropriate description is
given in terms of the supergravity solution of smeared (in the
eleventh circle direction) M2--branes \cite{imsy}. This result
also holds in presence of D6--branes, that also has a low energy
range described by smeared M2--branes \cite{ps}. It is natural to
expect that, if the D6--branes are wrapping a supersymmetric
cycle, the corresponding description will be given in terms of
smeared M2--branes transverse to some special holonomy manifold.
When we go further towards the IR, say $U < g^2_{YM}$, we expect
the smeared solution to be replaced (resolved) by a periodic array
of localized M2--branes along the eleventh circle. Closer enough
to the M2--branes, we should recover a conformal field theory. If
we KK reduce along the $y$--direction, we get: \beq ds^2_{10} =
\big[\,H(\rho)\,\big]^{-{1\over 2}}\,dx_{1,2}^2\,+\,
\big[\,H(\rho)\,\big]^{{1\over 2}}\,ds^2_7\,, ~~~~~ e^{\phi_D} =
\big[\,H(\rho)\,\big]^{{1\over 4}}\,\,, \label{cinueve} \eeq while
the 4-form field strength of D=11 becomes the RR 4-form $F^{(4)}$
of type IIA theory. It is clear that this solution represents a
D2--brane sitting at the tip of the $G_2$ holonomy manifold.
Notice, however, that the solution resulting from gauged
supergravity is the complete D2--brane solution. So, we should
reintroduce $l_p$ units everywhere and take $\rho$, $a$ and $l_p$
to zero such that $U \equiv a\rho/l_p^3$ and $L \equiv a^2/l_p^3$
are kept fixed. The asymptotic background gives the near horizon
limit of $N$ $D2$--branes transverse to the $G_2$ holonomy
manifold: \beq ds^2_{10} = l_s^2 \left( {U^{5 \over 2}\over
\sqrt{g_{YM}^2 N}} \,dx_{1,2}^2\,+\, {\sqrt{g_{YM}^2 N} \over
U^{{5\over 2}}} \,ds^2_7\, \right) \,, ~~~~~ e^{\phi_D} = \left(
{g_{YM}^{10} N \over U^5} \right)^{{1\over 4}}\,,
\label{cinuevenh} \eeq where $g_{YM}^2 \approx L$ is the three
dimensional coupling constant, $a l_s^2 = l_p^3$, and $N$ is the
number of $D2$--branes; the 4-form field strength $F^{(4)}$ is
unchanged.

In the UV we can trust the super Yang--Mills theory description.
In the case of a single $D2$--brane, it is an ${\cal N}=1$, $U(1)
\times U(1)$ gauge theory in $2+1$ dimensions with four complex
scalars $Q_i$, $\tilde Q_i$, $i=1,2$, and a vector multiplet whose
gauge field can be dualized to a compact scalar that would
parametrize the position of the $D2$--branes along the M--theory
circle. The vacuum moduli space is given by \beq |q_1|^2 + |q_2|^2
- |\tilde q_1|^2 - |\tilde q_2|^2 = L^2 ~, \label{vmoduli} \eeq
where $q_i$, $\tilde q_i$ are the scalar components of the
superfields $Q_i$, $\tilde Q_i$, which precisely provides an
algebraic--geometric description of the $G_2$ manifold.

In summary, we have briefly presented some aspects of the lower
dimensional gauged supergravity approach to the study of gravity
duals of D--branes wrapping cycles of special holonomy manifolds.
Let me end by mentioning that the twisting procedure can be
significantly generalized such that, for example, all $G_2$
metrics of cohomogeneity one with $SU(2) \times SU(2)$ isometry
can be obtained from Salam--Sezgin's theory \cite{epr2}.

\medskip
\medskip
I wish to thank Carlos N\'u\~nez, Kyungho Oh, \'Angel Paredes,
Alfonso Ramallo and Radu Tatar for delightful collaborations on
these subjects. I would also like to thank the organizers of the
Sakharov conference and the people of Moscow and St. Petersburg
for their kind hospitality. This work has been supported in part
by MCyT and FEDER under grant BFM2002-03881, by Xunta de Galicia,
by Fundaci\'on Antorchas and by Funda\c c\~ao para a Ci\^encia e a
Tecnologia under grants POCTI/1999/MAT/33943 and
SFRH/BPD/7185/2001.

\end{document}